# SU(2) LATTICE GAUGE THEORY WITH LOGARITHMIC ACTION: SCALING AND UNIVERSALITY[1]

Urs M. Heller

SCRI, Florida State University, Tallahassee, FL 32306-4052, USA.

## Abstract

We investigate a version of SU(2) lattice gauge theory with a logarithmic action. The model is found to exhibit confinement, contrary to previous claims in the literature. Comparing ratios of physical quantities, like $\sqrt{\sigma}/T_c$, we find that the model belongs to the same universality class as the standard SU(2) lattice gauge theory with Wilson action. Like the positive plaquette model, the model with logarithmic action has a monotonic $\beta$-function, without the famous dip exhibited by the Wilson action. Short distance dislocations affecting the definition of topology are slightly more suppressed than for the positive plaquette model.

---

# 1 Introduction

To investigate the non-perturbative properties of non-abelian gauge theories, by strong coupling expansions and more importantly numerical simulations, they are put on a space-time lattice following the approach of Wilson [1]. In Wilson's formulation it is then easy to show that the theory confines in the strong coupling region. Numerical simulations have produced a lot of evidence that confinement is not a mere lattice artefact but is a property likely to survive in the continuum limit [2, 3]. Nevertheless it has been suggested that the property of confinement in SU(N) lattice gauge theory is due to the use of elements of the compact gauge group, instead of elements of the Lie algebra [4], or due to the fact that the lattice action is bounded [5]. Grady proposed in Ref. [5] to use for an unbounded action the logarithm of the trace of the plaquette, and at the same time restrict the path integral to positive plaquettes like in the positive plaquette model (PPM) of Mack and Pietarinen [6, 7]. Grady claimed that he could see no sign of confinement in his model.

At $\beta = 0$ the PPM and Grady's logarithmic action model (LAM) are identical, since the action vanishes and the only restriction is to positive values of the plaquette. This restriction invalidates the usual proof of confinement at strong coupling. However it has recently been shown convincingly, albeit with numerical simulations, that the PPM confines [8] – at least at strong coupling. Furthermore Ref. [8] gave good evidence that the continuum limits of the PPM and SU(2) lattice gauge theory with standard Wilson action (SWA) are identical.

Therefore, Grady's conclusion that his LAM does not show confinement must be wrong at strong coupling. It is therefore interesting to re-investigate the model better, employing, for example, the recently developed enhancement techniques for the measurements of the heavy quark potential. In particular the scaling properties of the model need be investigated, to see whether the model with logarithmic action might lead to the same continuum limit as well, as universality arguments would suggest.

# 2 The model and perturbative considerations

Grady's LAM for gauge group SU(N) is given by [5]

$$\begin{aligned} S^{(G)} &= \sum_p S_p^{(G)} \\ S_p^{(G)} &= \begin{cases} -\beta \log\left[\frac{1}{2N}\text{Tr}(U_p + U_p^\dagger)\right] & \text{for} \quad \text{Tr}(U_p + U_p^\dagger) > 0 \\ \infty & \text{otherwise} \end{cases} \end{aligned} \quad . \quad (1)$$

Obviously, for configurations with $\text{Tr}(U_p + U_p^\dagger)$ close to zero for at least one plaquette, the action becomes unbounded.

Writing, as usual, $U_p = \exp(i\Psi_p)$, $\Psi_p = a^2 g F_{\mu\nu}(x) + \mathcal{O}(a^3)$, it easy to see that Grady's action has the correct naive continuum limit

$$S_p^{(G)} = \frac{\beta}{2N}\text{Tr}\Psi_p^2 + \mathcal{O}(\beta\psi_p^4) = a^4 \text{Tr} F_{\mu\nu}^2(x) + \mathcal{O}(a^6) \tag{2}$$

where we have used the usual relation, $\beta = 2N/g^2$. The model thus satisfies all the criteria to be a good lattice regularization of an SU(N) gauge theory. It is therefore important to check whether the model behaves the same as other regularizations also non-perturbatively, and leads to the same continuum limit, especially in light of claims to the contrary for the case of gauge group SU(2) [5]. But first we will briefly discuss results of lattice perturbation theory for the LAM that we shall need later on.

Most (lattice) perturbative results in the literature are known – or easily derivable from known results – for one-plaquette actions of the general form

$$S_p = \sum_R s_R(\beta) \left[1 - \frac{1}{2d_R}\text{Tr}_R(U_p + U_p^\dagger)\right] \tag{3}$$

where $R$ labels the representation of the group, $d_R$ is the dimension and $\text{Tr}_R$ the trace in the representation $R$. For the action eq. (3) to have the correct naive continuum limit the coefficients $s_R(\beta)$ have to satisfy

$$\sum_R \frac{s_R(\beta) T(R)}{d_R} = \frac{\beta}{2N} = \frac{1}{g^2}. \tag{4}$$

Since we are only interested in one-loop perturbative results, we will not attempt to rewrite Grady's logarithmic action entirely in the form eq. (3), but rather we will do so only up to $\mathcal{O}(\Psi_p^6)$, which is sufficient for our purposes. We find, for $N \geq 4$,

$$s_F^{(G)}(\beta) = 2\beta \;, \quad s_{R_b}^{(G)}(\beta) = -\frac{\beta(N+1)}{8N} \;, \quad s_{R_c}^{(G)}(\beta) = -\frac{\beta(N-1)}{8N} \;, \quad s_G^{(G)}(\beta) = -\frac{\beta(N^2-1)}{4N^2} \;. \tag{5}$$

Here $F$ and $G$ denote fundamental and adjoint representations of SU(N), and $R_b$ and $R_c$ the representations denoted by $(2, 0, 0, \cdots, 0)$ and $(1, 1, 0, \cdots, 0)$. For $N = 3$, $R_c \cong \overline{F}$ and we find

$$s_F^{(G)}(\beta) = \frac{23}{12}\beta \;, \quad s_{R_b}^{(G)}(\beta) = -\frac{1}{6}\beta \;, \quad s_G^{(G)}(\beta) = -\frac{2}{9}\beta \;. \tag{6}$$

For $N = 2$, finally, $R_b \cong G$ and $R_c \cong \mathbf{1}$, and we obtain

$$s_F^{(G)}(\beta) = 2\beta \;, \quad s_G^{(G)}(\beta) = -\frac{3}{8}\beta \;. \tag{7}$$

From this we see that Grady's action gives rise to negative adjoint couplings in addition to the fact that it excludes negative plaquettes. Both properties have been suggested to assure a smoother crossover from strong to weak coupling behavior [7, 9].

At weak coupling, we expand the action eq. (3) as

$$S_p = \frac{\beta}{2N}\frac{1}{2}\Psi_p^a\Psi_p^a - \frac{1}{4!}\sum_R \frac{s_R(\beta)}{d_R}\text{Tr}_R(\Psi_p^4) + \mathcal{O}(\Psi_p^6) \tag{8}$$

$$= \frac{\beta}{2N}\left\{\frac{1}{2}\Psi_p^a\Psi_p^a - \frac{1}{4!}\sum_R \frac{\hat{s}_R(\beta)}{d_R}\text{Tr}_R(\Psi_p^4) + \mathcal{O}(\Psi_p^6)\right\} \tag{9}$$

with $\hat{s}_R(\beta) = 2Ns_R(\beta)/\beta = g^2 s_R(\beta)$. The last form is useful to apply the results of Gonzales-Arroyo and Korthals-Altes [10]. Denoting by $\overline{\Lambda}$ the scale parameter of the action with $\hat{s}_R(\beta) = 0$ for all $R$, they find

$$\log\left(\frac{\overline{\Lambda}}{\Lambda^S}\right) = \frac{1}{2b_0}\left[\sum_R \frac{\hat{s}_R(\beta)C_2(R)T(R)}{4d_R} - \frac{N}{24}\right] \tag{10}$$

where $b_0$ is the one-loop coefficient of the $\beta$-function.

Using that $s_F^{(W)}(\beta) = \beta$ with all other $s_R$'s vanishing for the Wilson action and the above determined coefficients for Grady's logarithmic action we then easily obtain for the ratio of the $\Lambda$ parameters

$$\log\left(\frac{\Lambda^W}{\Lambda^G}\right) = -\frac{1}{2b_0}\frac{(N^2+1)}{8N}. \tag{11}$$

Therefore the couplings are related by

$$\frac{1}{g_W^2} = \frac{1}{g_G^2} + \frac{(N^2+1)}{8N} + \mathcal{O}(g^2) \tag{12}$$

or

$$\beta_W = \beta_G + \frac{(N^2+1)}{4} + \mathcal{O}\left(\frac{1}{\beta}\right). \tag{13}$$

By matching Creutz ratios of equal size at weak coupling, Grady has numerically estimated that $\beta_W - \beta_G = 1.1(2)$ for SU(2) [5]. This estimate agrees fairly well with the asymptotic value of $\frac{5}{4}$ from eq. (13). It is well possible that the shift between the two couplings grows even bigger towards the strong coupling region. Indeed we shall find this to be the case. Care is thus needed when choosing the couplings for numerical simulations with Grady's action.

Later on we will investigate the non-perturbative $\beta$-function using both blockspin MCRG methods as well as the "ratio method". To perturbatively improve the latter we need the one-loop expansions of Wilson loops. For the Wilson action they have been computed in Ref. [11]. It is easy to generalize that computation to actions of the form eq. (3). In the notation of [11] and with the additional definition

$$Z(r,t) = Y(r,t) + \overline{W}_2(r,t)^2 \tag{14}$$

with $Z(r,t) = \frac{1}{4}\left(1 - \frac{1}{V}\right)\overline{W}_2(r,t)$ on a symmetric lattice (we use small letters $r$ and $t$ here to denote the size of the Wilson loop), we obtain

$$\begin{aligned}\langle W(r,t)\rangle &= 1 - g^2\frac{(N^2-1)}{N}\overline{W}_2(r,t) - g^4(N^2-1)X(r,t) + g^4\frac{(2N^2-3)(N^2-1)}{6N^2}\overline{W}_2(r,t)^2 \\ &\quad - g^4\frac{(N^2-1)}{6N}\left[\sum_R 6g^2\frac{s_R(\beta)T(R)C_2(R)}{d_R} - N\right]Z(r,t).\end{aligned} \tag{15}$$

Note that only the coefficient of the last term has changed compared to the result for the standard Wilson action. Also, no new lattice sums need be computed. For Grady's logarithmic action this becomes

$$\begin{aligned}\langle W(r,t)\rangle &= 1 - g^2\frac{(N^2-1)}{N}\overline{W}_2(r,t) - g^4(N^2-1)X(r,t) + g^4\frac{(2N^2-3)(N^2-1)}{6N^2}\overline{W}_2(r,t)^2 \\ &\quad + g^4\frac{(N^2-1)(N^2+6)}{6N^2}Z(r,t).\end{aligned} \tag{16}$$

We remark that the dependence on the action in the perturbative expansion of Wilson loops at order $g^4$, (15), is exactly the same as in the $\Lambda$-parameter ratio, (10). It is then easy to convince oneself that the computation of the improved coupling $g_V^2$ [12] from the measured value of the average plaquette is the same for all one-plaquette actions. The same holds for the computation of the improved coupling $\beta_E$, suggested by Parisi [13], and its relation to $g_V^2$. The formulae are given in the appendix of Ref. [8].

## 3   The simulations

For most of the simulations, whose results will be described in subsequent sections, we have used a mixture of 4 "microcanonical overrelaxation" sweeps followed by one Metropolis sweep. The "microcanonical" step

$$U_l \to U_l' = W^\dagger U_l^\dagger W^\dagger \tag{17}$$

where $W$ is the "staple" projected onto the SU(2) group is for the logarithmic action no longer action preserving. It must, therefore, be accompanied by a Metropolis accept/reject step. For the SWA the microcanonical step is useful because it produces, typically, a rather "large" change in the link value and thereby helps reducing the auto correlation time. For the LAM it turns out that the acceptance rate becomes sufficiently large with increasing gauge coupling $\beta$ to help reduce the increase in auto correlation times, rendering this step useful for the LAM as well.

We don't know of a fast and easily implementable heat bath algorithm for the LAM. Therefore we resorted to a Metropolis algorithm with the size of the proposed random changes tuned to give an acceptance rate of about 50%.

We performed simulations on asymmetric lattices with $N_\tau = 2$ and 4 to study the deconfinement transition. In those simulations we measured the Polyakov line after every Metropolis sweep. In the simulations on symmetric lattices, used for the MCRG study and the measurement of the string tension, glueball masses and topological susceptibility, we typically measured after every 20 Metropolis sweeps. In this way, about the same amount of CPU time was spent on updating and on all the measurements. It also turned out that this spacing – recall that 4 overrelaxation sweeps were done before every Metropolis sweep – made the measurements essentially independent, *i.e.*, without detectable autocorrelations.

## 4  Deconfinement temperature

We have seen that, perturbatively, there is a shift of 1.25 between the coupling of Grady's LAM and the SWA coupling $\beta_W$. To see what this shift is non-perturbatively for stronger couplings we can measure some physical quantity and tune the couplings until this quantity is the same in both models (of course in the strong coupling region it is possible that the shift in the coupling is somewhat different for every physical observable considered). One of the easier physical observables is the deconfinement transition temperature when compared on lattices with equal temporal extent $N_\tau$. We have made this comparison for $N_\tau = 2$ and 4.

At $\beta = 0$ the LAM and the PPM become identical. The critical coupling for $N_\tau = 2$ in the PPM has recently been determined to be $\beta_c = 0.10(1)$ [8], slightly bigger than 0. We expect, therefore, that the transition for the LAM will also occur at a small positive $\beta$.

Like in the SU(2) theory with SWA, the deconfinement transition in the LAM is expected to be of second order, in the universality class of the 3 dimensional Ising model. We investigated the transition under this assumption by considering (Binder) cumulants built from powers of the Polyakov line, averaged over the space volume [14]. The cumulants on two different space-like volumes, $8^3$ and $12^3$, where compared and the critical coupling was determined from the crossing point. The value of the cumulant at the transition point is a universal quantity. We found it to be quite close to the Ising value thereby supporting the hypothesis that the deconfinement transition of the LAM is in the same universality class as the 3-d Ising model.

For the critical couplings we found $\beta_c = 0.0275(20)$ for $N_\tau = 2$, and $\beta_c = 0.517(3)$ for $N_\tau = 4$. For the model with SWA the critical couplings are $1.880(3)$ and $2.2986(6)$ respectively, giving shifts between the couplings of 1.853 and 1.782, considerably larger than the asymptotic weak coupling value of 1.25.

## 5  Heavy quark potential

Grady has claimed to have found no evidence for confinement in his LAM [5]. Though we have already shown that the model does have a finite temperature deconfinement transition, which moves to larger values of the coupling as $N_\tau$ is increased, we want to find more evidence for confinement and compute the string tension. Comparing ratios of the square root of the string tension to the deconfinement transition we can not only test scaling for the LAM but also check whether it leads to the same continuum limit as the model with SWA.

Grady attempted to extract the potential from ordinary, planar Wilson loops. At the lowest $\beta$ he considered, $\beta = 0.5$, which turned out to be near the critical coupling for the $N_\tau = 4$ deconfinement transition, the potential that he obtained had large errors and no unambiguous conclusions could be drawn. Using the MCRG results of section 8 we estimate that the next coupling Grady considered, $\beta = 1.0$, is around the critical coupling for $N_\tau = 12$, and hence a $15^4$ lattice is too small to reliably extract the zero temperature potential.

To improve on Grady's computation of the heavy quark potential at reasonable values of the coupling, we employ the recently developed signal improving technique [15] which replaces the links that make up the space-like segments of (time-like) Wilson loops with recursively constructed "smeared" links. Besides planar Wilson loops, with on-axis space-like segments, we also considered Wilson loops with off-axis space-like segments along the paths (1,1,0) and (2,1,0) and those related by the cubic symmetry.

Smearing of the space-like links gives a better overlap with the lowest state in the sum over contributing eigenvalues of the transfer matrix

$$W(\vec{R}, T) = \sum_i c_i \exp\{-V_i(\vec{R})\} \tag{18}$$

with $V(\vec{R}) = V_0(\vec{R})$, the heavy quark potential we would like to find. With a better overlap the effective potential, extracted from

$$V_T(\vec{R}) = \log\left(\frac{W(\vec{R}, T)}{W(\vec{R}, T+1)}\right) \tag{19}$$

will have an earlier plateau, where the statistical errors are still small.

We fit the potential, taken as the effective potential at a given $T$, to the usual form [16]

$$V(\vec{R}) = V_0 + \sigma R - \frac{e}{R} - f\left(G_L(\vec{R}) - \frac{1}{R}\right) \quad . \tag{20}$$

Here $G_L$ denotes the lattice Coulomb potential, which takes into account the lattice artefacts present at smaller distances; it helps in getting good fits that also include rather small distances. We used fully correlated fits with the covariance matrix estimated by a bootstrap method. In all cases, the best fit values obtained in this way did not differ significantly from those of naive, uncorrelated fits. To select one from the many possible fits, obtained when varying the range over which the fit is performed, we define a "quality" of the fit as the product of confidence level times the number of degrees of freedom divided by the relative error of the string tension. We selected the fits with the highest quality and list the results in Table 1. The $T$ in the third column gives the time separation from which the potential was determined as in eq. (19).

For the smallest $\beta$ value considered, corresponding to the critical coupling of a finite temperature system with $N_\tau = 2$ (see the previous section) rotational invariance of the potential is quite strongly violated. Thus we made fits over the potential along a principle axis, and along the (1,1,0) directions separately, and list the result in two rows of the Table. There was no sign of a Coulomb term – the string tension term completely dominates the potential – so we left it out from the fit.

For the next coupling, $\beta = 0.517$, the critical coupling of the $N_\tau = 4$ finite temperature transition, rotational invariance is already quite well restored. But, as an additional check we also made fits for the on-axis potential and the potential along direction (1,1,0) separately and list them in rows 2 and 3 for each $T$.

For the largest coupling considered, $\beta = 1.0$, we see in Table 1 big differences between the results on the $12^4$ and $16^4$ lattice and even on the latter one the fit values do not stabilize with increasing $T$. These are findings symptomatic of lattice sizes that are too small, as we have anticipated. And obviously, so was the lattice size, $15^4$, that Grady used.

| $\beta$ | $L$ | T | $V_0$ | $\sigma$ | e | f | range | $\chi^2/dof$ | CL |
|---|---|---|---|---|---|---|---|---|---|
| 0.0275 | 8 | 1 | 0.126( 6) | 0.675( 5) | | | 2.00 - 4.00 | 2.016/1 | 0.133 |
| | | 1 | 0.347(28) | 0.649(10) | | | 2.83 - 5.66 | 0.397/1 | 0.672 |
| | | 2 | 0.064( 2) | 0.698( 2) | | | 1.00 - 3.00 | 2.460/1 | 0.085 |
| | | 2 | 0.195(12) | 0.704( 8) | | | 1.41 - 4.24 | 0.296/1 | 0.744 |
| 0.517 | 12 | 3 | 0.588(10) | 0.1378(16) | 0.254(16) | 0.75(8) | 2.24 - 8.49 | 4.819/8 | 0.885 |
| | | 3 | 0.589(12) | 0.1374(22) | 0.271(14) | | 2.00 - 6.00 | 0.009/2 | 0.999 |
| | | 3 | 0.584( 7) | 0.1388(14) | 0.249( 7) | | 1.41 - 8.49 | 0.836/3 | 0.947 |
| | | 4 | 0.584(17) | 0.1388(28) | 0.249(25) | 0.77(11) | 2.24 - 8.49 | 5.120/8 | 0.854 |
| | | 4 | 0.543( 5) | 0.1461(14) | 0.218( 3) | | 1.00 - 6.00 | 2.306/3 | 0.595 |
| | | 4 | 0.587(12) | 0.1383(26) | 0.252(12) | | 1.41 - 8.49 | 2.817/3 | 0.465 |
| | | 5 | 0.588(12) | 0.1375(27) | 0.256(12) | 0.449(24) | 1.41 - 8.49 | 5.270/10 | 0.957 |
| | | 5 | 0.546( 9) | 0.1450(26) | 0.220( 6) | | 1.00 - 6.00 | 1.207/2 | 0.878 |
| | | 5 | 0.592(26) | 0.1370(58) | 0.257(25) | | 1.41 - 8.49 | 0.670/3 | 0.969 |
| | 16 | 3 | 0.511(24) | 0.1445(23) | 0.050(62) | 1.65(31) | 3.00 - 11.31 | 7.881/11 | 0.828 |
| | | 3 | 0.592( 6) | 0.1365(11) | 0.273( 7) | | 2.00 - 8.00 | 0.840/4 | 0.989 |
| | | 3 | 0.582( 4) | 0.1389( 7) | 0.246( 4) | | 1.41 - 11.31 | 1.282/5 | 0.990 |
| | | 4 | 0.586( 4) | 0.1375( 8) | 0.254( 4) | 0.452( 9) | 1.41 - 11.31 | 14.951/15 | 0.471 |
| | | 4 | 0.581(13) | 0.1389(24) | 0.263(15) | | 2.00 - 8.00 | 1.994/4 | 0.858 |
| | | 4 | 0.583( 7) | 0.1381(15) | 0.247( 7) | | 1.41 - 11.31 | 1.689/5 | 0.971 |
| | | 5 | 0.593( 6) | 0.1358(14) | 0.260( 6) | 0.459(12) | 1.41 - 11.31 | 11.208/15 | 0.839 |
| | | 5 | 0.632(27) | 0.1306(50) | 0.352(34) | | 2.00 - 8.00 | 4.041/4 | 0.426 |
| | | 5 | 0.582(14) | 0.1376(32) | 0.239(14) | | 1.41 - 11.31 | 1.182/5 | 0.993 |
| 0.75 | 12 | 3 | 0.561( 2) | 0.0521( 4) | 0.233( 2) | 0.272( 9) | 1.41 - 8.49 | 1.955/10 | 0.999 |
| | | 4 | 0.564( 3) | 0.0518( 5) | 0.235( 3) | 0.274(10) | 1.41 - 8.49 | 5.900/10 | 0.923 |
| | | 5 | 0.567( 3) | 0.0508( 6) | 0.239( 4) | 0.281(13) | 1.41 - 8.49 | 6.117/10 | 0.908 |
| | 16 | 3 | 0.558( 1) | 0.0525( 2) | 0.229( 1) | 0.270( 4) | 1.41 - 11.31 | 11.816/15 | 0.788 |
| | | 4 | 0.563( 1) | 0.0514( 2) | 0.233( 2) | 0.275( 6) | 1.41 - 11.31 | 10.032/15 | 0.915 |
| | | 5 | 0.563( 2) | 0.0514( 3) | 0.234( 2) | 0.277( 7) | 1.41 - 11.31 | 9.796/15 | 0.927 |
| | | 6 | 0.561( 2) | 0.0517( 4) | 0.232( 2) | 0.272( 8) | 1.41 - 11.31 | 7.464/15 | 0.990 |
| | | 7 | 0.563( 3) | 0.0512( 5) | 0.234( 3) | 0.285( 9) | 1.41 - 11.31 | 10.466/15 | 0.890 |
| 1.0 | 12 | 3 | 0.523( 2) | 0.0145( 3) | 0.227( 4) | 0.184(12) | 2.00 - 8.49 | 2.262/9 | 0.999 |
| | | 4 | 0.534( 3) | 0.0119( 4) | 0.241( 5) | 0.177(13) | 2.00 - 8.49 | 5.748/9 | 0.872 |
| | | 5 | 0.548( 8) | 0.0090( 4) | 0.258( 5) | 0.231(50) | 2.24 - 8.49 | 5.540/8 | 0.804 |
| | 16 | 4 | 0.517( 1) | 0.0171( 2) | 0.222( 2) | 0.177( 7) | 2.00 - 11.31 | 2.881/14 | 0.999 |
| | | 5 | 0.523( 1) | 0.0159( 2) | 0.230( 3) | 0.173( 8) | 2.00 - 11.31 | 6.799/14 | 0.990 |
| | | 6 | 0.529( 2) | 0.0147( 2) | 0.239( 3) | 0.220(32) | 2.24 - 11.31 | 9.680/13 | 0.821 |
| | | 7 | 0.533( 2) | 0.0138( 3) | 0.243( 4) | 0.238(36) | 2.24 - 9.90 | 8.879/12 | 0.815 |

Table 1: Fits to the potential approximants $V_T(R)$.

Having already determined the deconfinement transition on lattices with $N_\tau = 2$ and 4 we can now get the ratio $\sqrt{\sigma}/T_c$. We find 1.67(6) and 1.47(2), respectively, in excellent agreement with the corresponding numbers for the PPM [8] and the model with SWA. For the latter two models, $\sqrt{\sigma}/T_c$ was found to scale, *i.e.* be independent of the bare lattice coupling, starting at about the critical coupling for $N_\tau = 4$ [8, 17]. Since the ratios agree even before this scaling region, it is very likely that they will continue to do so, and that all three models have the same continuum limit.

For completeness we list in Table 2 the average plaquette values and in Table 3 the improved or effective couplings $g_V^2$ [12] and $\beta_E$ [13]. It has been noted in Ref. [8] that at equal physical conditions, the deconfinement transition for given $N_\tau$, the improved couplings of the SWA and the PPM are rather close. We can now see that this property persists for the LAM. Thus we can expect that asymptotic scaling with respect to one of the improved couplings will be quite similar in all models.

| $\beta$ | $L = 8$ | $L = 12$ | $L = 16$ |
|---|---|---|---|
| 0.0275 | 0.54814(4) | | |
| 0.517 | 0.42124(6) | 0.42128(2) | 0.42133(1) |
| 0.75 | 0.37620(6) | 0.37631(3) | 0.37639(1) |
| 1.0 | 0.33765(5) | 0.33767(2) | 0.33774(1) |

Table 2: Average plaquette values, $\langle 1 - \mathrm{Tr} U_p/2 \rangle$

| $\beta$ | $\beta_E$ | $g_E^2$ | $\beta_V$ | $g_V^2$ |
|---|---|---|---|---|
| 0.0275 | 1.36826 | 2.92341 | | |
| 0.517 | 1.78008 | 2.24709 | 1.03914 | 3.84934 |
| 0.75 | 1.99261 | 2.00741 | 1.27480 | 3.13776 |
| 1.0 | 2.22064 | 1.80128 | 1.51844 | 2.63428 |

Table 3: Effective couplings from the $L = 16$ plaquette values, except for $\beta = 0.0275$ where the $L = 8$ value was used.

# 6   Glueball measurements

Having found that $\sqrt{\sigma}/T_c$ at equal lattice spacing $a$, as determined from $T_c$, agrees between SWA, the PPM and now the LAM, we want to check whether this holds also for other ratios of physical quantities. We chose the glueball masses with quantum numbers $0^{++}$ and $2^{++}$ for this purpose. They can be obtained from correlations of plaquette operators built from the smeared spatial links that were already used for the computation of the potential. Experience with the PPM indicates that the operators built this way can not be further improved without including 6- or 8-link loops [8].

| $\beta$ | $L$ | $t$ | $m(0^{++})$ | $m(2^{++})$ I | $m(2^{++})$ II | $\sigma_{Pol}(L)$ |
|---|---|---|---|---|---|---|
| 0.517 | 12 | 1 | 1.58( 3) | 2.14( 5) | 2.17( 6) | 0.141( 3) |
|  |  | 2 | 1.61(14) | 2.54(69) | 2.09(42) | 0.117(14) |
|  | 16 | 1 | 1.60( 2) | 2.18( 5) | 2.15( 5) | 0.149( 4) |
|  |  | 2 | 1.34(12) | 2.32(43) | 2.06(40) | 0.189(61) |
| 0.75 | 12 | 1 | 0.98( 3) | 1.32( 3) | 1.35( 3) | 0.041( 1) |
|  |  | 2 | 0.93( 4) | 1.20( 8) | 1.24(12) | 0.038( 2) |
|  |  | 3 | 0.99(14) | 0.86(23) | 1.24(34) | 0.034( 2) |
|  | 16 | 1 | 1.01( 2) | 1.34( 3) | 1.32( 2) | 0.050( 1) |
|  |  | 2 | 0.85( 6) | 1.32(10) | 1.27(10) | 0.049( 2) |
|  |  | 3 | 0.74(10) | 1.27(41) | 0.71(24) | 0.047( 5) |

Table 4: Running glueball masses and string tension $\sigma_{Pol}(L)$.

The results for the running glueball masses are listed in Table 4. There we also give the string tension obtained from correlations of space-like Polyakov loops. This string tension, after the finite size correction by $\pi/(3L^2)$, agrees reasonably with the string tension extracted from the potential.

For the ratio $m(2^{++})/m(0^{++})$, when taking the running masses from $t = 1$, we obtain 1.36(4) and 1.32(4) for the two values of $\beta$. For the ratio from $t = 2$ we find 1.53(27) and 1.56(16). These results show good scaling, and they are, again, in good agreement to those of the PPM [8] and the SWA [18, 19]. Similarly for the ratio $m(0^{++})/\sqrt{\sigma}$ we find, with the mass extracted from $t = 1$, 4.36(7) and 4.46(10) for the two $\beta$ values studied, and with the mass extracted from $t = 2$ we get 3.65(34) and 3.75(26), in agreement with the results for the PPM and the SWA.

# 7 Topology

In the previous sections we have established that, with respect to the scaling behavior of ratios of physical observables, the LAM fares equally well as the model with SWA and the PPM. It is thus a viable alternate lattice gauge model. It is therefore interesting to investigate the behavior of topology in this model.

It is well known that for the standard SU(2) lattice gauge theory with SWA the unique assignment of an integer valued topological charge to an equilibrium configuration is hampered by lattice artefacts. Indeed, these short distance lattice artefacts, so called dislocations, are believed to dominate the topological susceptibility on ensembles of configurations produced in a numerical simulation. It has also been found that suppression of negative plaquettes [8] and an admixture of a negative adjoint coupling [20] help suppress, though not eliminate, the dislocations.

The LAM shares both these beneficial features: negative plaquettes are absent and the expansion of the action in irreducible representations reveals an effective negative adjoint coupling (see section 2). Hence there is a chance that measurements of the topological susceptibility are less ambiguous in the LAM. To test this we have measured the covariance between the 8 different computations of the topological charge with the geometric Philipps-Stone algorithm [21, 22], introduced in Ref. [8]

$$CV_{PS} = \frac{1}{28} \sum_{1 \leq i < j \leq 8} \frac{\langle (Q_i - \langle Q_i \rangle)(Q_j - \langle Q_j \rangle) \rangle}{\sqrt{\langle (Q_i - \langle Q_i \rangle)^2 \rangle \langle (Q_j - \langle Q_j \rangle)^2 \rangle}}. \quad (21)$$

To compare to the results obtained there, we also performed the computation at the critical coupling for the $N_\tau = 4$ deconfinement transition. On a $6^4$ lattice we found 0.545(13) as compared to 0.519(16) in the PPM and 0.404(15) for the SWA, while on an $8^4$ lattice the results are 0.570(15), 0.531(15) and 0.397(13). We see that the outcome for the LAM is by a slight margin better, but it is still far from the value 1 that a unique charge assignment would give.

We have to conclude that in the LAM some dislocations are still present and that it hence does not give a solution to the problem of obtaining reliable numbers for the topological susceptibility on the lattice. In order to compare with results from the PPM and with SWA, we employ the same methods, with all their caveats, that have been used in those models: the cooling method [23] to eliminate the short distance fluctuations plus measurement of the naive, non-integer topological charge with a "clover" definition for $F_{\mu\nu}$ on the lattice.

| $\beta$ | $L$ | Hot, PS | 10c, N1 | 10c, N2 | 15c, PS | 15c, N1 | 15c, N2 | 20c, N1 | 20c, N2 |
|---|---|---|---|---|---|---|---|---|---|
| 0.517 | 8 | 17.3(8) | 3.17(14) | 5.53(21) | 4.76(21) | 3.04(13) | 5.00(21) | 2.91(13) | 4.45(19) |
| | 12 | 17.0(6) | 3.03(9) | 4.17(11) | 4.63(14) | 2.97(9) | 4.09(11) | 2.87(9) | 3.94(10) |
| | 16 | | 2.98(9) | 3.57(10) | | 2.96(9) | 3.57(10) | 2.90(9) | 3.48(10) |
| 0.75 | 8 | 1.95(11) | 0.42(3) | 0.68(5) | 0.59(5) | 0.40(3) | 0.60(5) | 0.38(3) | 0.57(4) |
| | 12 | 2.30(11) | 0.72(3) | 1.11(5) | 0.96(4) | 0.72(3) | 1.04(4) | 0.71(3) | 0.99(4) |
| | 16 | | 0.79(4) | 1.09(4) | | 0.81(4) | 1.09(4) | 0.81(4) | 1.07(4) |

Table 5: Topological susceptibility $\times 10^4$ and cooling. "PS" denotes the Phillips-Stone charge, "N1" the naive (non-integer) charge and "N2" the naive charge but rounded away from zero.

We also assign an integer charge to each configuration by rounding this non-integer charge to the nearest integer away from zero ("ceiling" of the absolute value). We computed the topological susceptibility with both definitions of the naive charge. They are listed in Table 5. For the $8^4$ and $12^4$ configurations we also used the geometric Philipps-Stone algorithm [21] for the topological charge measurement. We did these measurements on the "'hot" configurations, and after 15 coolings sweeps. The latter helps in estimating the systematic uncertainties that result from use of a naive definition of the topological charge, even on cooled configurations. The large drop of the susceptibility between hot and cooled configurations is another indication that the susceptibility on the hot configurations is dominated by short distance dislocations also in the LAM.

In order to compare our results with those of the PPM we use the same ad hoc definition as Ref. [8]: we take the susceptibility from the naive charge rounded away from zero (N2) after 15 cooling steps on the $16^4$ lattices. Then we obtain for $\chi_t^{1/4}/\sqrt{\sigma}$ 0.373(3) and 0.449(4) for the two $\beta$-values listed in Table 5. These values agree well with those of the PPM and SWA, when compared at equal lattice spacing as obtained for example from the string tension. We conclude that once the short distance fluctuations, including the damaging dislocations, are "cooled away" all three actions lead to the same results also with respect to topology. Whether the topological susceptibility computed with the cooling method, on the other hand, has any physical meaning is a question outside the scope of this paper.

# 8 MCRG results

So far we have investigated the scaling behavior of the LAM and found good agreement both with the PPM and with the SWA. The latter two differ, though, quite dramatically in their asymptotic, *i.e.*, two-loop, scaling behavior. The Wilson action produces the famous dip in the "step $\beta$-function", $\Delta\beta$, while suppression of negative plaquettes in the PPM removed this dip [8]. On the other hand, the approach to asymptotic scaling was found to be, if anything, slower in the PPM then for SWA. The LAM, as already mentioned, also completely suppresses negative plaquettes, and in addition induces effectively a negative adjoint coupling. Therefore we expect no dip in the step $\beta$-function.

To verify this expectation and to check the approach to asymptotic scaling, we computed the step $\beta$-function, $\Delta\beta$, for a change of scale by a factor 2, with two different Monte Carlo Renormalization Group (MCRG) methods, a real space RG with Swendsen blocking transformation [24, 9], and the so called "ratio method" [25].

For the real space RG, we made, at each $\beta = \beta_L$ on the larger lattice, a few trial runs on lattices smaller by a factor 2, comparing $8^4$ with $4^4$, to narrow down the range of the optimization parameter $p$ of the blocking scheme. Then we made runs with two values of $p$ that appeared to bracket the optimal value, comparing $12^4$ with $6^4$ and $16^4$ with $8^4$ lattices. For the matching we used two runs on the smaller lattices with slightly different $\beta_S$ and interpolated linearly to find the matching coupling for each of four observables, the plaquette and the three different shaped loops of length 6, built from blocked links at the appropriate blocking level. The results, interpolated to the optimal $p$ value, are listed in Table 6. The errors quoted include the statistical errors, determined by jackknife, as well as systematic errors describing the spread of $\Delta\beta = \beta_L - \beta_S$ obtained from the different observables. The columns with the label "extr." give the results after extrapolating from blocking levels $n$ and $n-1$ to an infinite number of blockings, taking into account only the leading irrelevant eigenoperator of the blocking RG transformation with eigenvalue $1/4$ [9],

$$\Delta\beta^{(\infty)} = \frac{1}{3}\left[4\Delta\beta^{(n)} - \Delta\beta^{(n-1)}\right]. \tag{22}$$

To implement the ratio method we measured planar Wilson loops with the signal improved by the method of Parisi, Petronzio and Rapuano [26]. Since for the LAM we do not have a closed formula to compute the improved links, we calculated them with typically 20 Metropolis hits. To increase the number of Wilson loops from which we can build ratios, especially for the smaller lattices, we measured Wilson loops up to distances $L/2 + 2$. Since the finite size effects on the large and small lattices to be matched are the same this is acceptable for the ratio method.

| $\beta_l$ | $L$ | 2 | 2, extr. | 3 | 3, extr. |
|---|---|---|---|---|---|
| 0.517 | 8 | 0.469(27) | 0.467(27) | | |
| | 12 | 0.478(16) | 0.478(13) | | |
| | 16 | 0.477(17) | 0.478(14) | 0.488(13) | 0.487(14) |
| 0.75 | 8 | 0.352( 8) | 0.350( 6) | | |
| | 12 | 0.346( 5) | 0.350( 7) | | |
| | 16 | 0.346( 5) | 0.350( 7) | 0.340( 2) | 0.338( 2) |
| 1.0 | 8 | 0.350( 7) | 0.355( 3) | | |
| | 12 | 0.334( 4) | 0.334( 6) | | |
| | 16 | 0.331( 4) | 0.330( 6) | 0.316( 4) | 0.311( 2) |

Table 6: Blocking MCRG results: the columns with label "$n$" contain the results from matching after $n$ and $n-1$ blocking steps. The columns with label "$n$, extr." contain the results from extrapolating $\Delta\beta$ to an infinite number of blocking steps according to eq. (22).

Each Wilson loop ratio considered provides a $\Delta\beta$, again inferred from two $\beta_S$ values on the smaller lattices by interpolation, with an error determined by jackknife. The results from "bare" ratios are afflicted by lattice artefacts, especially if the ratio contains small Wilson loops. These artefacts can be corrected for in perturbation theory by building appropriate linear combinations that give the correct result in perturbation theory at tree or one-loop level [25]. We have already described the differences in the perturbative expansion of Wilson loops between the logarithmic and standard Wilson action in section 2 and thus the relevant perturbative results are easily obtained.

Typically the improvement shows up in a smaller variance of the results from many different ratios. The results listed in Table 7 represent the average $\Delta\beta$ over all ratios satisfying certain "cuts". The cuts are minimal area of Wilson loops, area difference in numerator and denominator of a ratio and maximal statistical error from a ratio. The errors quoted in Table 7 are the variances over the ratios that passed all cuts.

The results from the two different RG methods, listed in Tables 6 and 7, agree within errors. The difference between the critical couplings for the deconfinement transition for $N_\tau = 4$ and 2 gives another $\Delta\beta$ to compare with: $\Delta\beta(\beta = 0.517) = 0.489(4)$ in good agreement with the MCRG results. The results are shown in Figure 1. As anticipated, the step $\beta$-function for the LAM has no dip, and the approach to asymptotic scaling seems

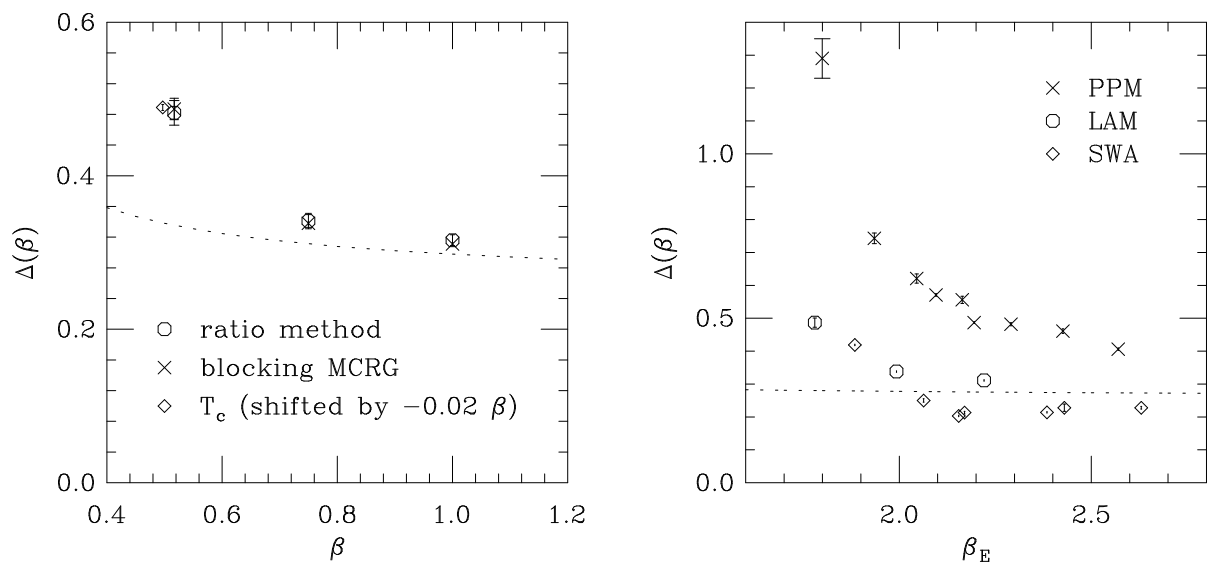

Figure 1: The step $\beta$-function versus the bare lattice coupling $\beta$ for the LAM (left) and versus the effective coupling $\beta_E$ (right). There we compare to blocking results for the PPM [8] and to results from $T_c$ and blocking for the SWA from Refs. [17, 27].

| $\beta_l$ | $L$ | basic | tree-level | 1-loop | $R_{max}$ | $\delta_{max}$ | $A_{min}$ | $\Delta A_{min}$ |
|---|---|---|---|---|---|---|---|---|
| 0.517 | 8 | 0.50(5) | 0.473( 4) | 0.474( 8) | 3 | 0.050 | 1 | none |
| | 12 | 0.50(4) | 0.468(15) | — | 4 | 0.100 | 2 | 0 |
| | | 0.49(5) | 0.464(19) | 0.465(26) | 3 | 0.100 | 1 | none |
| | 16 | 0.54(3) | 0.497(26) | 0.482(16) | 5 | 0.100 | 2 | 1 |
| | | 0.55(3) | 0.493(27) | 0.491(20) | 4 | 0.100 | 2 | 0 |
| 0.75 | 8 | 0.35(4) | 0.330(14) | 0.338(21) | 3 | 0.020 | 1 | none |
| | 12 | 0.39(3) | 0.351(23) | 0.326( 5) | 4 | 0.020 | 2 | 0 |
| | | 0.35(5) | 0.326(28) | 0.332(30) | 3 | 0.020 | 1 | none |
| | 16 | 0.36(3) | 0.344(13) | 0.341(10) | 5 | 0.020 | 2 | 1 |
| | | 0.38(3) | 0.358(20) | 0.340( 4) | 4 | 0.020 | 2 | 0 |
| 1.0 | 8 | 0.36(6) | 0.333(18) | 0.339(35) | 3 | 0.030 | 1 | none |
| | 12 | 0.37(3) | 0.342(15) | 0.326( 5) | 4 | 0.010 | 2 | 0 |
| | | 0.34(5) | 0.297(26) | 0.322(21) | 3 | 0.010 | 1 | none |
| | 16 | 0.34(3) | 0.321(12) | 0.316( 7) | 5 | 0.010 | 2 | 1 |
| | | 0.36(4) | 0.332(15) | 0.318( 3) | 4 | 0.010 | 2 | 0 |

Table 7: Ratio method MCRG results. The two lines for sizes 12 and 16 represent different cuts on the Wilson loops and ratios considered, as described in the text.

considerably faster than for the PPM.

## 9  Conclusion

A careful investigation of the LAM of Grady [5], with unbounded action, showed, contrary to Grady's claim, confinement at strong coupling. Computation of the critical coupling for the finite temperature deconfinement transition, the potential and from it the string tension, and of the low-lying glueball spectrum showed that dimensionless ratios of physical quantities, when compared at equal lattice spacing, agree with those of the standard SU(2) lattice gauge theory with Wilson action and with the PPM. We conclude that all three are in the same universality class and have identical continuum limits.

As far as topology is concerned, dislocations seem to be more suppressed in the LAM than for the SWA or in the PPM, but there are still dislocations present that impede the unambiguous computation of the topological susceptibility.

Like for the PPM, the step $\beta$-function for the LAM shows no dip and approaches the asymptotic two-loop value monotonically. The approach to the two-loop value seems fastest for the LAM. Therefore one might argue that the LAM should be preferable over the PPM and the SWA.

This work was supported in part by the DOE under grants # DE-FG05-85ER250000 and # DE-FG05-92ER40742. The computations have been carried out on the cluster of Alpha workstations at SCRI. The author would like to thank J. Fingberg and V. Mitrjushkin for many discussions on the methods employed in this work, and M. Müller-Preussker for discussions on the subject of topology on the lattice and for providing an implementation of the Phillips–Stone algorithm.